# The Hydrodynamic Interaction in Polymer Solutions Simulated with Dissipative Particle Dynamics

**Wenhua Jiang[1], Jianhua Huang[1,2], Yongmei Wang[1]∗, and Mohamed Laradji[3]**

[1] Chemistry Department, The University of Memphis, Memphis, TN 38152-3390;
[2] Chemistry Department, Zhejiang Sci-Tech University, Hangzhou, 310018, China
[3] Physics Department, The University of Memphis, Memphis, TN 37152-3390; and MEMPHYS-Center for Biomembrane Physics, University of Southern Denmark, DK-5230, Denmark
∗ Corresponding author: ywang@memphis.edu

## Abstract

We analyzed extensively the dynamics of polymer chains in solutions simulated with dissipative particle dynamics (DPD), with a special focus on the potential influence of a low Schmidt number of a typical DPD fluid on the simulated polymer dynamics. It has been argued that a low Schmidt number in a DPD fluid can lead to underdevelopment of the hydrodynamic interaction in polymer solutions. Our analyses reveal that equilibrium polymer dynamics in dilute solution, under a typical DPD simulation conditions, obey the Zimm model very well. With a further reduction in the Schmidt number, a deviation from the Zimm model to the Rouse model is observed. This implies that the hydrodynamic interaction between monomers is reasonably developed under typical conditions of a DPD simulation. Only when the Schmidt number is further reduced, the hydrodynamic interaction within the chains becomes underdeveloped. The screening of the hydrodynamic interaction and the excluded volume interaction as the polymer volume fraction is increased are well reproduced by the DPD simulations. The use of soft interaction between polymer beads and a low Schmidt number do not produce noticeable problems for the simulated dynamics at high concentrations, except that the entanglement effect which is not captured in the simulations.

# 1. Introduction

The dynamics of polymer chains in solution is a fundamental subject in polymer physics.[1,2] Understanding the dynamics of polymer chains is key to predicting properties of polymer solutions such as diffusion coefficients, viscosity, sedimentation coefficient, and various rheological properties. The current state of knowledge of polymer dynamics can be found in several monographs that very nicely summarize the development in the field.[1,3] The first successful model of polymer dynamics was developed by Rouse.[4] In this model, a polymer chain is modeled as a string of beads connected by springs. The only interaction taken into account within the Rouse model is that between consecutive beads via the springs, and the hydrodynamic interaction between beads is ignored. The Rouse model leads to the prediction that a chain diffusion coefficient $D \sim N^{-1}$, and a chain relaxation time $\tau \sim NR^{-2}$, where $N$ is the number of beads in the chain and $R$ is the size of the chain. In dilute solution, however, the hydrodynamic interaction between the chain monomers is important and cannot be ignored. The Zimm[5] model, an extension of the Rouse approach, takes into account hydrodynamic interactions through the use of Navier-Stokes equation for describing the hydrodynamics of an incompressible solvent in the viscous regime. Within the Zimm approach, the relaxation of the solvent velocity field is assumed to occur at a much faster rate than the motion of polymer beads. The hydrodynamic interaction between the polymer beads is then accounted for by using the Oseen tensor. The Zimm model predicts the dependencies of the chain diffusion coefficient and the chain relaxation time on the chain size, given by $D \sim R^{-1}$, and $\tau \sim R^3$. Using the scaling dependence of $R$ on $N$ in a good solvent, $R \sim N^\nu$, where $\nu \approx 3/(d+2)$ is the Flory's exponent and $d$ is the spatial dimension, the Zimm model predicts that $D \sim N^{-\nu}$ and $\tau \sim N^{3\nu}$. Experiments indeed confirm that polymer dynamics in dilute solutions is best described by the Zimm model. For example, the Brownian motion of single, long, and fluorescently labeled DNA molecules, was observed to obey the Zimm model in a 10mMNaCl buffer solution[6], i.e., the chain diffusivity scales with its molecular weight as $D \sim N^{-3/5}$.

Computer simulations are very useful to gain a microscopic understanding of the dynamics of polymer solutions. For instance, Brownian dynamic simulations of bead-spring and bead-rod models of polymer chains in implicit solvents have been used to shed lights on chain stretching in shear flow.[7,8] Such simulations were able to provide results that are in good agreement with DNA single-molecule experiments in shear flow,[9] even though the hydrodynamic interaction between the beads were not incorporated in the simulations. On the other hand, comparison of the numerical results obtained from molecular dynamics simulations of polymer chains in implicit solvents with results obtained from dielectric spectroscopy experiments[10] reveals the importance of hydrodynamic interactions in dilute solution. In that study, the solvent molecules are not explicitly included in the simulation, and therefore hydrodynamic effects are not accounted for. As a result, these simulations yielded a Rouse-type scaling for the normal mode relaxation time, which fail to agree with dielectric spectroscopy experiments in dilute solutions. However, simulation results are found to be in accord with experiments in solutions with concentration $\rho > 8\rho^*$, where $\rho^*$ is the crossover concentration from the dilute to the semidilute regime.[1,2] The disagreement between the simulation and dielectric spectroscopy experiment in dilute solutions is ascribed to the neglect of the hydrodynamic interaction in the simulation. At high concentrations, however, the hydrodynamic interaction is screened, leading to good agreement between the simulation and experiments.

There is a strong interest in developing simulation approaches that are able to correctly take into account hydrodynamic interactions between polymer beads. Such simulation methods, once established, can be used to study a variety of interesting problems. Molecular dynamics (MD) simulations of polymer chains in explicit solvent have been used to simulate polymer dynamics in solutions.[11-14] In this approach, the hydrodynamic interactions result naturally from the intermolecular interactions. However, the fact that in such simulations most of the system is composed of solvent particles combined with a very small time step due to the use of Lennard-Jones interaction make this approach computationally very expensive. More efficient and fast algorithms than MD simulations are therefore desired. Jendrejack *et al*[15] recently developed a self-consistent Brownian dynamics approach with fluctuating hydrodynamic interactions. In this approach, the solvent is treated implicitly as a continuum medium, and the hydrodynamic interaction between polymer beads is approximated by Stokeslet formalism.[16-18] Ahlrichs and Dunweg[19] proposed to use a bead-spring model for the polymer chains coupled to a lattice Boltzmann description of the solvent to simulate polymer dynamics in solution. This approach has recently been used by Usta *et al.* to study equilibrium and non-equilibrium dynamics of polymer solutions confined in channels.[19-21]

Dissipative Particle Dynamics (DPD) is a novel mesoscale computational tool that has attracted a good level of popularity for studying soft materials during recent years. DPD was initially proposed by Hoogenbrugge and Koelman[22] and modified to its present form by Español.[23,24] DPD is reminiscent of molecular dynamics in that it is an explicit particle-based method but which uses soft conservative interactions and pairwise dissipative and random forces. The dissipative and random forces act collectively as a thermostat while locally conserving momentum. The local momentum conservation is necessary for a correct description of long-range hydrodynamics.[25] DPD has been used by us and others to investigate a variety of soft matter problems such as spinodal decomposition, microphase separation of block copolymers,[26] nanocomposites,[27] the dynamics of lipid bilayers[28], and solvent flow through polymer brushes.[29,30] Recently, however some concern has been raised regarding the capability of DPD in accounting for hydrodynamic interactions in dilute polymer solutions.[18] This concern was first expressed by Groot and Warren[31] who indicated that in typical DPD simulations, the Schmidt number is rather small, on the order of 1. The Schmidt number is defined as $Sc \equiv \mu_k/D_s$, where $\mu_k$ is the kinematic viscosity and $D_s$ is the diffusivity of a solvent particle. The Schmidt number of a typical solvent such as water is on the order of 1000. A low Schmidt number implies that a particle's diffusion is comparable or smaller than momentum transport. Technically, Zimm's theory is valid when the Schmidt number is much higher than 1. Therefore, it has been argued that if the Schmidt number within DPD simulations is on the order of 1, the hydrodynamic interactions are still developing on the timescale of bead motion. Hence, the dynamics of polymer chain simulated within DPD may not have full hydrodynamic interactions.[18] Furthermore, the use of soft repulsive interactions in DPD may allow for chain crossing, violating topological constraints.[32] Since DPD has successfully been used in a variety of problems, it is important to evaluate the extent of these concerns. One of the ways to evaluate these is through the comparison of the dynamics of polymer chains in solution, as revealed in DPD simulations, against the known theoretical predictions. While there have been several earlier reports that examined the dynamics of polymer solutions with DPD simulations,[33-35] these studies did not provide a clear evidence to the validity of DPD simulations for polymer solution dynamics.

In the present article, we report on extensive and systematic DPD simulations of polymer solutions in bulk conditions with concentrations ranging from dilute to semidilute and compared simulation results against the theoretical predictions. We discuss our results in the light of hydrodynamic interaction in polymer solution. In Sec. 2, we present the model and computational approach used in our simulations. In Sec. 3, our results are presented and discussed in detail. Finally, a summary of the essential results and conclusion are presented in Sec. 4.

## 2. Model, Numerical Approach and Simulation Details

In the dissipative particle dynamics (DPD) approach, solvent particles are coarse-grained into fluid elements, thereafter called *dpd*-particles.[31] These *dpd*-particles interact with each other via pairwise conservative, random and dissipative forces that locally conserve momentum leading to a correct hydrodynamic description.[36] The net force experienced by *i*th particle due to other particles within some radius cutoff, $r_c$, is given by

$$\mathbf{f}_i = \sum_{j \neq i} \left( \mathbf{F}_{ij}^{(C)} + \mathbf{F}_{ij}^{(D)} + \mathbf{F}_{ij}^{(R)} \right), \tag{1}$$

where $\mathbf{F}_{ij}^{(C)}$, $\mathbf{F}_{ij}^{(D)}$, $\mathbf{F}_{ij}^{(R)}$ are the conservative, dissipative, and random forces, respectively. DPD is differentiated from MD by the use of soft repulsive conservative interactions. This allows for time steps in DPD that are much larger than those in typical MD simulations. The three pairwise forces are given by

$$\mathbf{F}_{ij}^{(C)} = a_{ij} w(r_{ij}) \hat{\mathbf{r}}_{ij}, \tag{2}$$

$$\mathbf{F}_{ij}^{(D)} = -\gamma w^2(r_{ij})(\hat{\mathbf{r}}_{ij} \cdot \mathbf{v}_{ij}) \hat{\mathbf{r}}_{ij}, \tag{3}$$

and

$$\mathbf{F}_{ij}^{(R)} = \frac{\sigma}{(\Delta t)^{1/2}} w(r_{ij}) \theta_{ij} \hat{\mathbf{r}}_{ij}, \tag{4}$$

where $\mathbf{r}_{ij} = \mathbf{r}_i - \mathbf{r}_j$, $\hat{\mathbf{r}}_{ij} = \mathbf{r}_{ij} / |\mathbf{r}_{ij}|$, $\Delta t$ is the time step, and $\mathbf{v}_{ij} = \mathbf{v}_i - \mathbf{v}_j$. $\theta_{ij}$ is a symmetric random noise with zero mean and unit variance and is uncorrelated for different degrees of freedom and different times. The softness of the interaction is determined by the weight function, $w(r)$, for which we adopt the commonly used choice, $w(r) = 1 - r/r_c$ for $r \leq r_c$, and $w(r) = 0$ for $r > r_c$, where $r_c$ is the cutoff radius and will be used to set all other length scales in the problem. The interaction amplitude $a_{ij}$ determines the strength of the repulsive interaction and depends on the types of the particles *i* and *j*. Eqs. (3) and (4) are interrelated through the fluctuation-dissipation theorem at some temperature $T$, leading to $\sigma^2 = 2\gamma k_B T$, where $k_B$ is Boltzmann's constant. The motion of particle $i$ is governed by Hamilton's equations,

$$d\mathbf{r}_i = \mathbf{v}_i dt,  \qquad (5)$$

and

$$d\mathbf{v}_i = \frac{1}{m_i}\mathbf{f}_i dt,  \qquad (6)$$

where $m_i$ is the mass of particle $i$. For simplicity, in this work we consider the case where all particles have the same mass corresponding to $m$.

In the present study, we consider fully flexible polymer chains in a good solvent condition. Their integrity is ensured via an additional intra-chain interaction for which we use the finitely-extensible non-linear elastic (FENE) potential between consecutive beads.[37,38]

$$U_{\text{intra}}(r_{i,i+1}) = \begin{cases} -\dfrac{k_F}{2}(r_{\max} - r_{\text{eq}})^2 \ln\left[1 - \left(\dfrac{r_{i,i+1} - r_{\text{eq}}}{r_{\max} - r_{\text{eq}}}\right)^2\right] & \text{for } r_{i,i+1} < r_{\max}, \\ \infty & \text{for } r_{i,i+1} \geq r_{\max}, \end{cases}  \qquad (7)$$

where $k_F$ is an elastic coefficient, $r_{\text{eq}}$ is the preferred bond length at equilibrium, and $r_{\max}$ is the maximum bond length. We have used

$$k_F = 40.0(\varepsilon/r_c^2), \quad r_{\text{eq}} = 0.7 r_c, \quad \text{and} \quad r_{\max} = 2.0 r_c.  \qquad (8)$$

In Eqs. (8) $\varepsilon$ sets the energy scale. In the present study, we have two types of particles, corresponding to solvent particles and polymer beads. The interaction strengths between the three types of pairs correspond to $a_{ss} = a_{pp} = 25\varepsilon/r_c$ and $a_{sp}$ is varied. In our simulations, we choose a fluid density $\rho = 3.0 r_c^{-3}$, and a dissipation coefficient $\sigma = 3.0(\varepsilon^3 m/r_c^2)^{1/4}$. All our simulations were performed on cubic boxes, with a linear size $L$ ranging between $15 r_c$ and $30 r_c$, and subject to periodic boundary conditions along the three directions. We consider polymer solutions with volume fractions ranging from the dilute regime, $\phi < 0.025$ to the concentrated regime, $\phi = 0.7$. The equations of motion, (5) and (6), were integrated using the velocity-Verlet algorithm with a time step $\Delta t = 0.04 \tau_0$, where the DPD time scale $\tau_0 = (mr_c^2/\varepsilon)^{1/2}$.

The equilibrium properties of the polymer solution are characterized by the mean square radius of gyration of the polymer chains and their end-to-end distance. The equilibrium dynamics is characterized by the mean-square displacement of the center-of-mass of a chain,

$$(\Delta \mathbf{R}_{CM}(t))^2 = \langle (\mathbf{R}_{CM}(t+t_0) - \mathbf{R}_{CM}(t_0))^2 \rangle, \tag{9}$$

where $\mathbf{R}_{CM}$ is the position of the center of mass defined as $\mathbf{R}_{CM} = (1/N)\sum_{i=1}^{N}\mathbf{R}_i$ and $\mathbf{R}_i$ is the coordinate of monomer $i$. Eq. (9) allows for the extraction of single chain diffusion coefficient, $D$, from the equation

$$(\Delta \mathbf{R}_{CM}(t))^2 = 6Dt. \tag{10}$$

The brackets in Eqs. (9) denote both an average over all chains in the solution and over many starting times, $t_0$, at equilibrium conditions. In the extraction of the diffusion coefficient, the fit of the mean-square displacement of the center-of-mass to a linear form, Eq. (10), is restricted to the interval $1\tau$ to $2\tau$, where $\tau$ is the longest relaxation time of the chain, as determined from analyses of the autocorrelation function. We also calculate the autocorrelation function of the end-to-end vector, defined as

$$C(t) = \frac{\langle \mathbf{R}_{1N}(t+t_0)\cdot \mathbf{R}_{1N}(t_0) \rangle}{\langle \mathbf{R}_{1N}^2 \rangle}, \tag{11}$$

where $\mathbf{R}_{1N}$ is the end-to-end vector of a polymer chain. Here too, the brackets indicate an average over all chains and time origins, $t_0$. The relaxation time is obtained by fitting the autocorrelation function of the end-to-end vector to an exponential form,

$$C(t) = C_0 \exp(-t/\tau). \tag{12}$$

The fitting is performed over the time domain when $C(t)$ decays from 1.0 to about 0.1. In addition, we also calculated the relaxation of internal modes of the chains. A polymer chain composed of $N$ beads is characterized by $N$ relaxation modes, with index $p = 1, 2, \ldots, N$. The relaxation time, $\tau_p$, of the $p$th mode represents the relaxation time of a sub-polymer chain containing $N/p$ consecutive monomers. In order to determine the $p$th relaxation time, we introduce the set of Rouse coordinates $\{\mathbf{X}_p; p = 1, \ldots, N\}$ defined according to

$$\mathbf{X}_p(t) = \frac{1}{N} \sum_{j=1}^{N} \cos[p\pi(j-1/2)/N] \, \mathbf{R}_j(t), \tag{13}$$

where $\mathbf{R}_j(t)$ is the position of the $j$th bead of a polymer chain. From Eq. (13), the mean square of the Rouse coordinate, $\mathbf{X}_p$, is given by

$$\langle \mathbf{X}_p^2 \rangle = -\frac{1}{2N^2} \sum_{i,j=1}^{N} \cos[p\pi(i-1/2)/N]\cos[p\pi(j-1/2)/N] \langle (\mathbf{R}_i - \mathbf{R}_j)^2 \rangle. \tag{14}$$

It has been recently shown that for long chains[19]

$$\langle \mathbf{X}_p^2 \rangle = \frac{N^{2\nu}}{p^{2\nu+1}} f_\nu(p), \tag{15}$$

where $f_\nu(p)$ is slightly greater than one for self-avoiding walk chains and depends very weakly on $p$. We will define

$$R_p = \left(p \langle \mathbf{X}_p^2 \rangle \right)^{1/2} \tag{16}$$

as the amplitude of each Rouse mode. Neglecting the function $f_\nu(p)$, then $R_p$ scales as $(N/p)^\nu$, same as $R_g$ of a chain. The values of $R_p$ were computed directly from the simulations. The normalized autocorrelation function, $C_p(t)$, is defined as

$$C_p(t) = \frac{\langle \mathbf{X}_p(t+t_0) \cdot \mathbf{X}_p(t_0) \rangle - \langle \mathbf{X}_p \rangle^2}{\langle \mathbf{X}_p^2 \rangle - \langle \mathbf{X}_p \rangle^2}. \tag{17}$$

The relaxation time $\tau_p$ of mode $p$ is then obtained by measuring the decay of the autocorrelation function assuming an exponential form, $C_p(t) = C_{p0} \exp(-t/\tau_p)$, over the range from 1.0 to 0.1.

## 3. Results and Discussion

### 3.1 Dilute athermal solution

In this subsection, we present simulation data obtained for dilute polymer solutions with a polymer volume fraction of polymer, $\phi < 0.06$ in an athermal solvent, i.e with $a_{sp} = a_{pp} = a_{sp} = 25\varepsilon/r_c$. The number of monomers per chain is varied between $N=10$ to 100. In order to investigate the effect of finite size of the simulation box, we considered four values of the box linear size, corresponding to $L = 15$, 20, 25, and $30 r_c$. All static and dynamic quantities are evaluated based on a single trajectory with length up to 200 to $400\tau$, where $\tau$ is the estimated longest relaxation time of a chain. In some case, for long chains and in large simulation boxes, the trajectory length is reduced to $100\tau$ due to limitation in CPU time. The estimations of error bars on static quantities are based on the fluctuations of the mean and the mean-square of the quantity, divided by the number of statistically uncorrelated samples. The error bars on dynamic quantities were estimated from differences between three components. The typical value of the error is less than 1% for static quantities, but may reach up to 10% for dynamic quantities.

**3.1.A Static Properties**: Results of the mean square radius of gyration, $\langle R_{g0}^2 \rangle$, and the mean square end-to-end distance, $\langle R_{1N}^2 \rangle$, are shown in Fig. 1. We found that the box size has very small effect on the static properties. Values $\langle R_{g0}^2 \rangle$ and $\langle R_{1N}^2 \rangle$ increased slightly as the box size increases. The power-law fit of $\langle R_{g0}^2 \rangle$ with the number of bonds, $\langle R_{g0}^2 \rangle \sim r_c^2 (N-1)^{2\nu}$, yields $2\nu$ = 1.154±0.008, 1.154±0.009, 1.16±0.01, and 1.17±0.01, for linear box size $L$=15, 20, 25, and $30 r_c$, respectively. Similar respective exponent from a power-law fit of $\langle R_{1N}^2 \rangle$ yields $2\nu$ = 1.185±0.003, 1.178±0.004, 1.19±0.02, and 1.21±0.01, which are slightly larger than those for $\langle R_{g0}^2 \rangle$. The error bars on the exponent increases as $L$ increases, partly because the trajectory length decreases with increasing box size. We must note that the difference in the exponents however is small between different box sizes. Taking all the data in Fig. 1 for the power-law fit, we find that $\langle R_{g0}^2 \rangle = (0.103 \pm 0.008) r_c^2 (N-1)^{1.16 \pm 0.02}$ and $\langle R_{1N}^2 \rangle = (0.57 \pm 0.010) r_c^2 (N-1)^{1.19 \pm 0.02}$. This implies a Flory exponent, $\nu$=0.58 ±0.01 from the radius of gyration and $\nu$=0.595±0.01 from the end-to-end distance. In the remainder of discussion, we will assume $\nu$=0.59 ±0.01 in our model. These results are in good agreement with previous DPD simulations of dilute solutions of polymer chains with harmonic intra-chain interactions.[33,35,39] These data shows that the simulated chains possess an effective excluded volume interaction, and therefore there is no need to introduce additional repulsive interactions between polymer beads in order to enforce excluded volume interaction.[40]

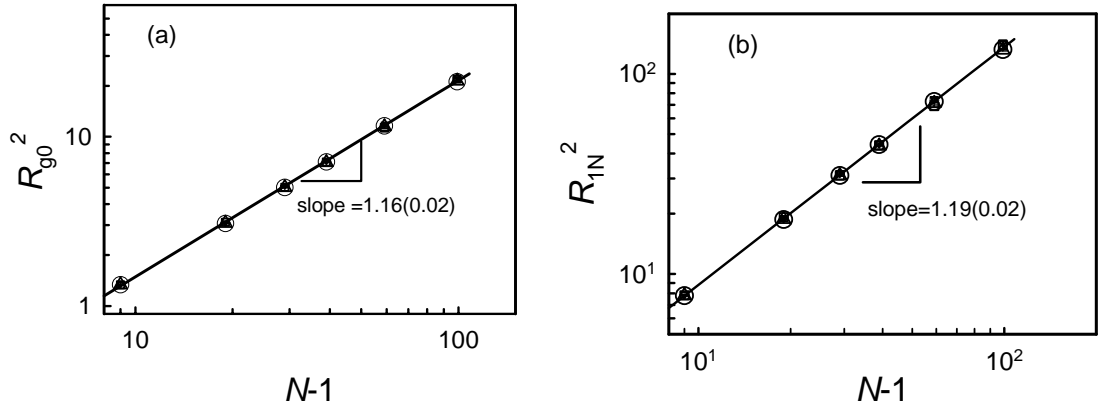

**Fig. 1** The mean square radius of gyration (a) and the mean square end-end distance (b) for polymer in dilute bulk solution. The symbols represent results obtained from four periodic systems with linear size $L=30r_c$ (O), $L=25r_c$ (□), $L=20r_c$ (○) and $L=15r_c$ (Δ). The solid lines correspond to the power-law fits of simulation data. All error bars are smaller than the symbols.

**3.1.B Diffusion coefficients:** The diffusion coefficient obtained from the mean-square displacement of the center of mass, $(\Delta R_{CM}(t))^2$, using Eq. (10), exhibits a strong system size dependence, as depicted in Fig. 2. We should note that since the polymer volume fraction is very low, the numerically calculated diffusion coefficient corresponds to a chain center-of-mass self-diffusion coefficient. This figure shows that polymers diffusivity decreases as the box size decreases and that this effect is more significant for longer polymer chains. Fig. 2 also shows an apparent decrease in the scaling exponent of the diffusivity with $N$ from 0.81 to 0.67, as the system size is increased from 15 to $30r_c$.

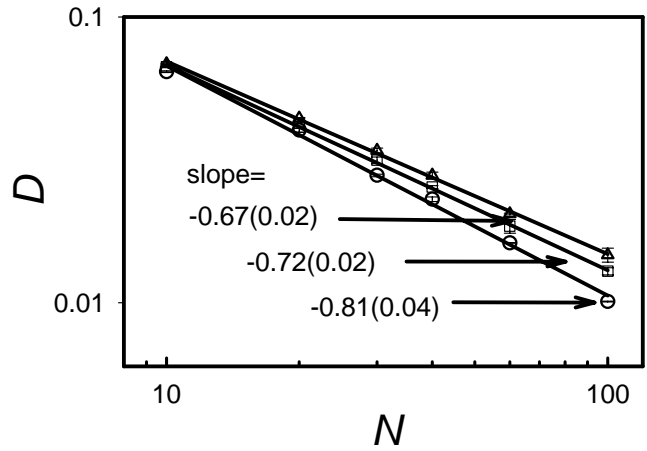

**Fig. 2** Log-log plot of the diffusion coefficient, $D$, as a function of chain length, $N$, determined from the DPD simulations in three periodic systems with linear size: $L=30r_c$ (Δ), $L=20r_c$ (□) and $L=15r_c$ (○). The solid lines represent fits of simulation data with power laws.

Previous studies empirically suggested that the static and dynamic properties in the thermodynamic limit and at equilibrium should be achieved when $L/R_g > 5$.[33,35,39] However, we still found a strong box size effect for the diffusivity even when $L/R_g > 5$. This strong finite size effect on the diffusion coefficient has been discussed in several articles.[11,13,41] In order to investigate this issue even further, we extrapolated our data for the diffusivity obtained at finite system sizes to the thermodynamic limit, since the finite box size corrections are proportional to $1/L$,[13] as clearly demonstrated by Fig. 3. The diffusion coefficient in the thermodynamically large systems is obtained from the intercept, with the y-axis, of the fits of the data with a linear form.

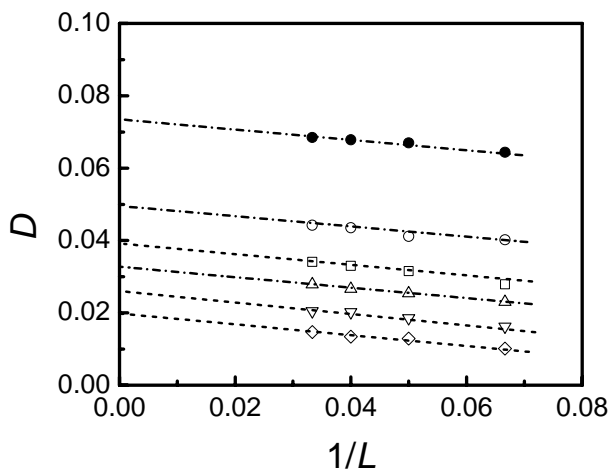

**Fig. 3** The extrapolation of the diffusion coefficient, $D$, simulated in four boxes ($L$=15, 20, 25 and $30r_c$) to the thermodynamic limit. The symbols (from top to bottom) correspond to $N$=10(●), $N$= 20 (○), $N$=30 (□), $N$= 40 (Δ), $N$=60 (∇), and $N$=100(◊), respectively. The dash lines are fits of the data with a linear form.

The extrapolated diffusion coefficient in the thermodynamic limit, $D_\infty$, is shown as a function of $N$ in Fig. 4. This figure shows that $D_\infty = (0.284 \pm 0.023)(r_c^2/\tau_0)N^{-(0.59\pm 0.02)}$. This result is in agreement with the Zimm model for a chain in a good solvent, i.e., $D_\infty \propto N^{-\upsilon}$, with $\nu = 0.59$.

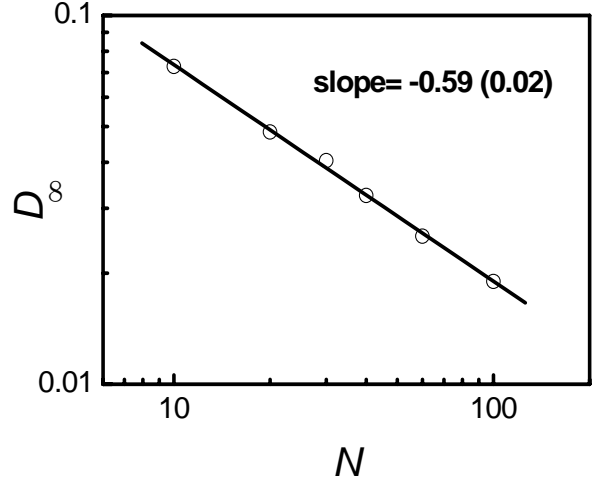

**Fig. 4** Log-log plot of the diffusion coefficient, $D_\infty$, (obtained from Fig. 3 ) as a function of chain length, $N$. The solid line represents a linear fit of the data.

**3.1.C Relaxation times**: The longest relaxation time, $\tau$, computed from the autocorrelation function of the end-to-end vector, Eq. (11), shows a very weak dependence on the system size, mostly within errors, as demonstrated by Fig. 5. The exponent in the power-law dependence between $\tau$ and $N$ has an error of about 10%. Our best estimation for the relaxation time is $\tau = (0.13 \pm 0.01)\tau_0 N^{(1.81\pm 0.06)}$. In the very dilute regime, if the chains obey the Zimm model, we expect $\tau \sim N^{3\nu} \sim N^{1.77}$. On the other hand, if the chains obey the Rouse model, then $\tau \sim N^{1+2\nu} \sim N^{2.18}$ is expected. Clearly, our data shown in Fig. 5(a) is much more in accord with the Zimm model than with the Rouse model. A better test of whether the data is in accord with the Zimm model or the Rouse model would be to plot $\tau$ against the radius of gyration $R_g$ determined in the simulation. We recall that the Zimm model predicts $\tau \sim R_g^3$ regardless of the value of $\nu$. However, the Rouse model would predict $\tau \sim R_g^{2+1/\nu}$ with the exponent $2 + 1/\nu \approx 3.69 \pm 0.03$ taking $\nu \approx 0.59 \pm 0.01$. Fig. 5(b) clearly shows that the scaling of $\tau$ with $R_g$ obeys predictions from the Zimm model.

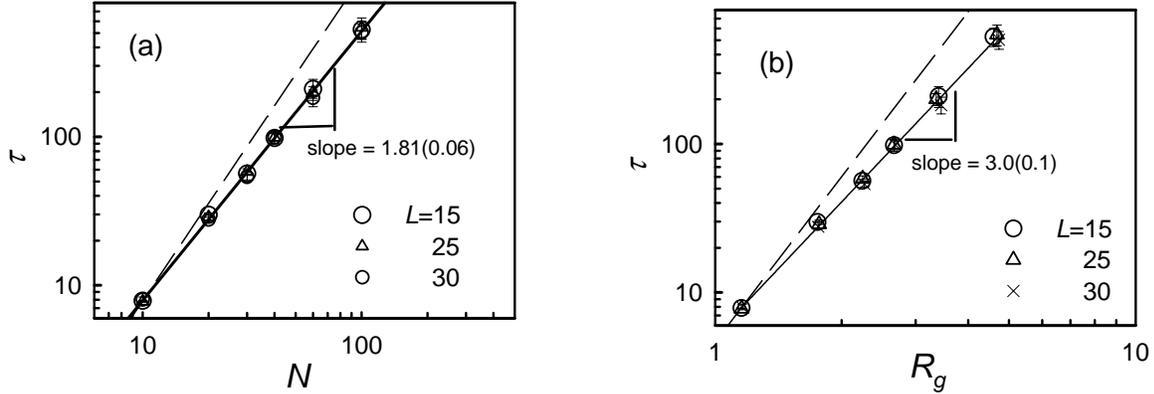

**Fig. 5** Log-log plot of the relaxation time, $\tau$, computed from end-to-end autocorrelation, as a function of (a) polymer chain length, $N$, and (b) the radius of gyration, $R_g$ for three system sizes, $L=30\ r_c$ (×), $L=25\ r_c$ (Δ), and $L=15\ r_c$ (○). The solid lines are linear fits to determine the scaling exponent. The long dashed lines show the power law as expected from the Rouse model.

**3.1.D Relaxation mode analysis:** A more stringent test of chain dynamics, than that shown in the previous subsection, is to study the dependence of relaxation time $\tau_p$ of mode $p$ and on the amplitude of each mode $R_p$. The relaxation time, $\tau_p$, of the $p$th mode of a polymer chain of length $N$ represents the relaxation of a sub-chain of length $N/p$. If hydrodynamic interactions were absent, then the relaxation time $\tau_p$ would scale with $p$ according to the Rouse model, i.e., $\tau_p = \tau_m (N/p)^{1+2\nu}$, where $\tau_m$ is the relaxation time of a monomer. However, if hydrodynamic interactions are fully present within a subchain, then $\tau_p$ would scale with $p$ according to the Zimm model, i.e., $\tau_p = \tau_m (N/p)^{3\nu}$. Alternatively, if we use the amplitude of motion, $R_p$ defined in Eq. (16), then we obtain the dependence of $\tau_p \sim R_p^3$ for the Zimm model, and $\tau_p \sim R_p^{2+1/\nu}$ for the Rouse model. Fig. 6(a) shows the scaling plot of $\tau_p$ with $N/p$ for a polymer chain with $N = 20, 30, 60$, and $100$. The scaling dependence of $\tau_p$ on $N/p$ is well observed and the scaling exponent is $1.84 \pm 0.03$, slightly above the expected $3\nu \approx 1.77 \pm 0.03$ from the Zimm model, but well below $1 + 2\nu \approx 2.18 \pm 0.02$ as expected from the Rouse model. We note that a recent more refined treatment of the Zimm model predicts that $\tau_p = \tau_m (N/p)^{3\nu} r_\nu(p)$, where $r_\nu(p)$ is close to one and has a weak dependence of on $p$.[14,19] This correction term slightly raises the observed scaling exponent. When $\tau_p$ is plotted against $R_p$, we found that $\tau_p \sim R_p^{3.05 \pm 0.06}$, which suggests a very good agreement with the predictions from the Zimm model. We believe that the analysis of the relaxation time is better assessed when $\tau_p$ is analyzed with respect to the amplitude of the Rouse mode $p$, i.e $R_p$, since the Zimm model would predict an exponent, 3.0, independent of the value of $\nu$.

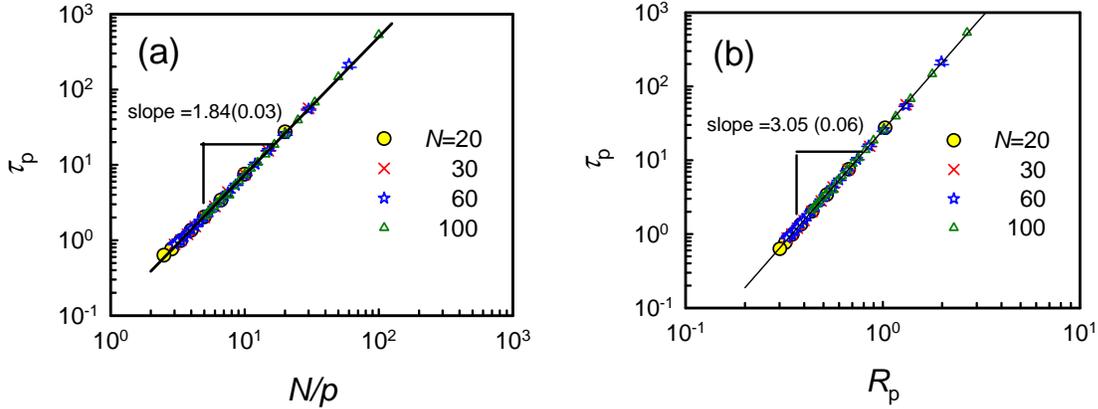

**Fig. 6**: (a) Log-log plot of relaxation time $\tau_p$ versus $N/p$ for $N$ =20, 30, 60 and 100. (b) Log-log plot of $\tau_p$ versus $R_p$ for the same systems shown in (a). The solid line is the best fit of the data to determine the scaling exponent.

**3.1.E Subdiffusive monomers motion:** The mean-square displacement of the monomers on the chains are known to exhibit a subdiffusive regime for time scales shorter than the longest relaxation time $\tau$ of a single chain. In order to investigate whether the DPD model of polymer chains does indeed exhibit this behavior, we calculated the mean-square displacement of the five innermost monomers on the chain.

$$g_1(t) = \frac{1}{5} \sum_{i=N/2-2}^{N/2+2} \left\langle [\mathbf{R}_i(t+t_0) - \mathbf{R}_i(t_0)]^2 \right\rangle. \tag{18}$$

This choice of central monomers within the chain is motivated by the fact that the subdiffusive motion is sensitive to the position of beads along the chain.[42]

Theoretical arguments based on the Zimm model predict that the means square displacement obeys the following scaling behavior[1]

$$g_1(t) = \begin{cases} t^1 & \text{for} \quad t < \tau_m \\ t^{2/3} & \text{for} \quad \tau_m < t < \tau \\ t^1 & \text{for} \quad t > \tau \end{cases}. \tag{19}$$

If the chain dynamics obeys the Rouse model, however, then for the time regime $\tau_m < t < \tau$, $g_1(t) \sim t^{1/2}$ for ideal chains, and $g_1(t) \sim t^{0.54}$ for self-avoiding chains. In Fig. 7, the mean square displacement, $g_1(t)$, is shown for three different chain lengths corresponding to $N$ = 30, 60 and 100. A subdiffusive regime is clearly observed. The exponent obtained by fitting the data in the subdiffusive regime (which we define as the interval where $1.0 < g_1(t) < R_g^2$) is found to be 0.75, 0.72 and 0.70 for $N$=30, 60 and 100, respectively; all are slightly larger than 2/3, but well above the value of the exponent as predicted from the Rouse model for self-avoiding chains, i.e. 0.54. At very short time scales or long time scales, Fig. 5 shows that $g_1(t) \sim t^1$, in accord with Eq. (19). According to the scaling theory, the exponent of $g_1(t)$ expected in the subdiffusive regime is a direct result of the dependence of $\tau_p$ on $R_p$. The slight discrepancy between our numerical value of the exponent of $g_1(t)$ with respect to time in the subdiffusive regime is puzzling, since we have found that $\tau_p \sim R_p^{3.05}$, in very good agreement with the Zimm model. Nevertheless, we must note that the exponent decreases as the chain length increases. The subdiffusive motion of monomers arises from the chain connectivity. We should note that the subdiffusive motion of monomers which are close to the two ends of a chain, exhibit an exponent that is even larger. We therefore believe that the slight deviation of the exponent from that of the Zimm model is due to contribution from end monomers, whose effect is diminished as the chain length is increased. In the next section, we will show that as the polymer volume fraction is further increased, leading to an increased screening of hydrodynamic interactions, the subdiffusive exponent approaches its value predicted from the Rouse model, which is 0.54.

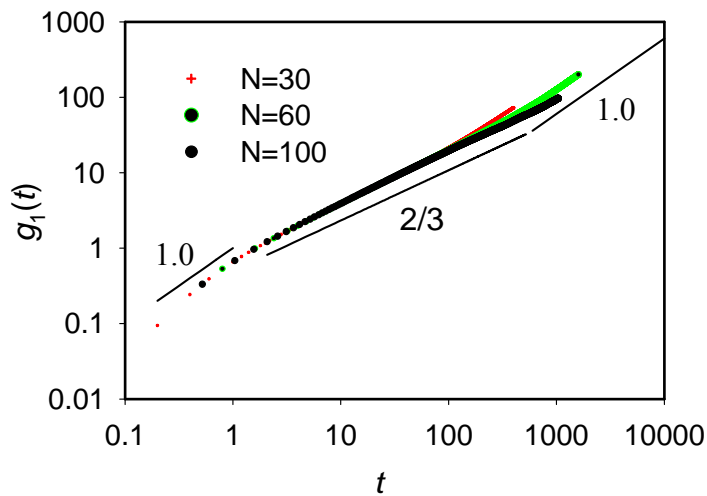

**Fig. 7**: Log-log plot of monomer mean-square displacement $g_1(t)$ versus $t$. The solid lines indicate the slope 1.0 at early time (t<$\tau_m$), 2/3 according to the Zimm model in the subdiffusive regime, and 1.0 at later time ($t>\tau$).

## 3.2 Semidilute athermal solutions

In this section, we present results obtained from the simulation of athermal polymer solutions with a volume fraction systematically increased from 0.03 to 0.7 and for chain length $N$ =25, 50 and 100. The trajectories lengths in these simulations are typically about $100\tau$ for short chains and about 50τ for $N$=100. As the polymer volume fraction is increased above the overlap volume fraction, $\phi^*$, two things occur. First, the excluded volume interaction becomes screened and the static properties of ideal chains are recovered. Second, the hydrodynamic interactions also become screened and the Rouse model is recovered. The two effects are believed to occur concurrently. Therefore, there exists only one blob length, $\xi$, within which both the hydrodynamic interaction and the excluded volume interactions are present. For length scales beyond the blob size, however, both the hydrodynamic interaction and the excluded volume interactions are screened. To our knowledge, these two screening effects have not been properly investigated within the DPD approach.

**3.2.A Static Properties:** We first present results on the behavior of the static properties of the chains with increasing polymer volume fraction. As the polymer volume fraction increases, the excluded volume interaction becomes screened and the swelling of the chains is reduced. According to de Genne's scaling theory, the size of a linear polymer chain in a semidilute solution in a good solvent is given by[1]:

$$\frac{R_g}{R_{g0}} \sim \left(\frac{\phi}{\phi^*}\right)^{-(2\nu-1)/(6\nu-2)} \quad (18)$$

where $R_{g0}$ is the radius of gyration in the dilute limit and $\phi^*$ is the overlap volume fraction calculated according to $\phi^* = N/\left[(4/3)\pi\rho R_{g0}^3\right]$. In our case, $\phi^* = $ 0.24, 0.13 and 0.08 for polymer solutions with $N$ =25, 50 and 100, respectively. Taking $\nu$=0.59, the exponent in Eq. (18) becomes $(2\nu-1)/(6\nu-2) = 0.12$. Fig. 8(a) depicts the dependence of the radius of gyration on polymer volume fraction as obtained from our DPD simulations. The scaling dependence of radius gyration with $\phi/\phi^*$ is well observed and the observed exponent agrees well with de Gennes prediction. In Fig. 8(b) DPD simulation data for $R_g/R_{g0}$ is shown together with data obtained from lattice Monte Carlo simulations.[43] The two sets of data overlap very well, except for the first three points in DPD data. These three points correspond to the lowest volume fractions investigated with DPD for each chain length. In the current DPD simulations, the lowest volume fraction investigated is $\phi$=0.03, which may not be dilute enough especially for $N$ =100 for which $\phi^* \approx 0.08$. The model-independent scaling shown in Fig. 8(b) suggests that the screening of excluded volume interaction in semidilute solution is well captured by the DPD simulation. In particular, the soft interactions employed in DPD simulations do not pose problems as far as the static properties are concerned.

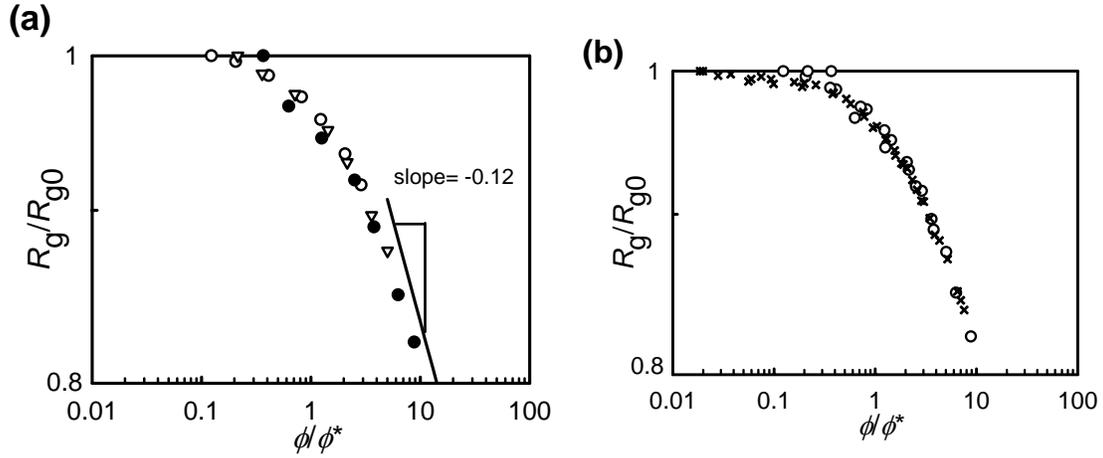

**Fig. 8** Dependence of the polymer chains radius of gyration on their volume fraction; (a) Data obtained from DPD simulations of polymer chains, $N=25$ (○), 50 (▽) and 100 (●), and (b) Overlap of data obtained from DPD simulations (○) and from lattice Monte Carlo simulations (×). The straight line represents the theoretical scaling predictions in a semidilute solution.

The screening of excluded volume interaction can also be detected from the analysis of Flory's exponent, $\nu$, with respect to the polymer volume fraction. Fig. 9 shows that the exponent $\nu$ decreases from 0.59 to 0.515, as the polymer volume fraction is increased up to 0.7, a clear indication of the screening of excluded volume interactions. From DPD simulations, Spenley verified earlier that $\nu=0.50$ in the melt condition.[35]

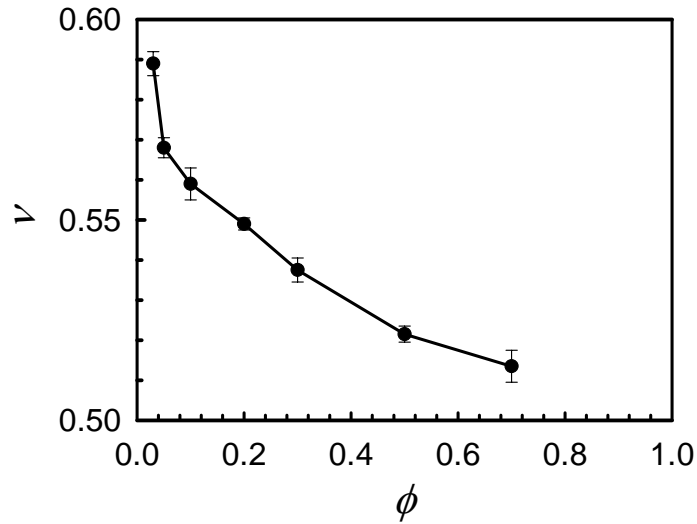

**Fig. 9:** The Flory's exponent, $\nu$, extracted from scaling plot of $R_g \sim (N-1)^\nu$, for different polymer volume fractions $\phi$, in an athermal solvent. The solid line is only a guide for the eyes.

**3.2.B Relaxation times**: The volume fraction dependence of polymer relaxation times in solution has been investigated theoretically[44,45] and experimentally.[46] According to de Genne's blob theory, the longest relaxation time of a polymer chain in a semidilute unentangled solution is given by[1]

$$\frac{\tau}{\tau^{(0)}} \sim \left(\frac{\phi}{\phi^*}\right)^{(2-3\nu)/(3\nu-1)} \tag{19}$$

where $\tau^{(0)}$ is the relaxation time in the dilute limit. In a good solvent, taking $\nu \approx 0.59$, the exponent $(2-3\nu)/(3\nu-1) \approx 0.30$. Fig. 10 shows the relaxation time deduced from the end-to-end vector autocorrelation function, Eq. (11). Again, the scaling law is well observed and the expected scaling exponent is reached at high volume fractions. However, one should note that the increase of $\tau$ at high volume fractions increases by a factor slightly lower than 2.0 for the highest volume fraction investigated. Unfortunately, the scaling theory does not provide us the prefactor in Eq. (19). If one assumes the prefactor is 1.0, then the estimated $\tau$ according to Eq. (19) is in the right order of magnitude as the obtained simulation data. According to Eq. (19), short chains will have an overall smaller increase in $\tau$ than long chains, since $\phi^*$ decreases with increasing $N$. Earlier theoretical work by Muthukumar[44,45] and experimental studies by Patel *et al*[46] suggested that the relaxation time in dilute solution should increase exponentially with the polymer volume fraction. Our data however does not exhibit an exponential increase; rather, the relaxation time, after an initial increase, crosses over to the predicted scaling regime given by Eq. (19). It is possible that in the crossover regime the relaxation time increases exponentially with the polymer volume fraction. However, we are unable to make this assessment at the moment since our chains are relatively short. Another earlier study, using MD simulations,[10] with chain lengths comparable to ours, reported an exponential increase of relaxation time with volume fraction and a maximum factor of increase close to 4.0. In that study, however, the solvent molecules are not explicitly included. As a result, the drag observed in the dilute regime may be artificially small and therefore the increase in drag as the polymer volume fraction increase could be artificially enhanced. On the other hand, we note that DPD simulations with models similar to ours do not capture the entanglement effect in the melt condition with long chains,[32,34] whereas MD simulations of polymer melts with long chains do seem to capture entanglement effects.[42] The use of a soft interaction potential between polymer beads in DPD simulations is believed to be the main cause of failing to reproduce the entanglement effect. This could have also contributed to an overall small magnitude of increase in $\tau$ observed at high volume fractions in our current study.

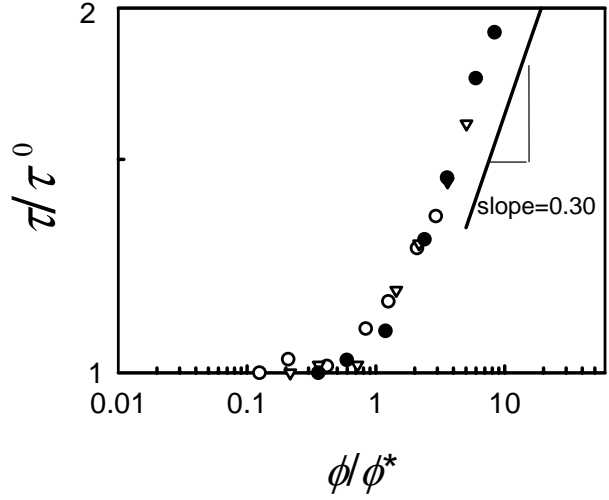

**Fig. 10:** Log-log plot of relaxation time from end-to-end vector autocorrelation function versus volume fraction of polymers with $N=25$ (○), $N=50$ (∇) and $N=100$ (●), in a good solvent. The straight line shows the theoretical scaling prediction in a semidilute solution.

**3.2.C Subdiffusive monomers motion:** Earlier, we discussed the subdiffusive motion of the monomers in the dilute regime. There, we verified that this regime is characterized by a monomer mean square displacement, $g_1(t) \sim t^{2/3}$ for intermediate time scales $\tau_m < t < \tau$, in accordance with predictions based on the Zimm model. In the absence of hydrodynamic interactions, however, the Rouse model predicts that $g_1(t) \sim t^{1/2}$. As the polymer volume fraction is increased, hydrodynamic interactions become increasingly screened. Therefore, one expects a crossover from the Zimm-like subdiffusive monomer motion to a Rouse-like subdiffusive motion. In Fig. 11, the mean-square displacement of the monomers, $g_1(t)$, is shown for various values of polymer volume fraction and chain length. At low volume fractions, $(\phi = 0.05, N = 50)$, we observe a Zimm regime, $g_1(t) \sim t^{2/3}$, at times $t < \tau$, followed by the usual long time diffusive regime at late times (characterized by $g_1(t) \sim t$ for $t > \tau$). As the polymer volume fraction is increased, we observe an initial Zimm regime, followed by a crossover to a Rouse regime, $g_1(t) \sim t^{1/2}$, at intermediate times. Eventually the system crosses over to the diffusive regime at later times. These findings are in agreement with a recent study by Ahlrichs *et al*[16] using a hybrid model that combines a lattice-Boltzmann model for the solvent with a MD description for the polymer chains. There it was pointed out that the hydrodynamic screening in semidilute solution should be viewed as a time-dependent as well as a spatial-dependent process.

At short time, the Zimm behavior dominates regardless of the spatial distances. At longer time, the screening of hydrodynamic interaction leads to the Rouse behavior. Additionally, we also note that if entanglement effects are present, a reptation motion in concentrated polymer solutions, with a power law $g_1(t) \sim t^{1/4}$ should be observed. Clearly, such regime is absent in our data. There is no entanglement effect in these simulations. This could be partly because the chain length investigated is not long enough. Reptation is anticipated for very long chains, possibly much longer than those considered in our DPD simulations. There is also the possibility that due to the use of soft interaction in DPD, chains crossing reduce reptation kinetics.

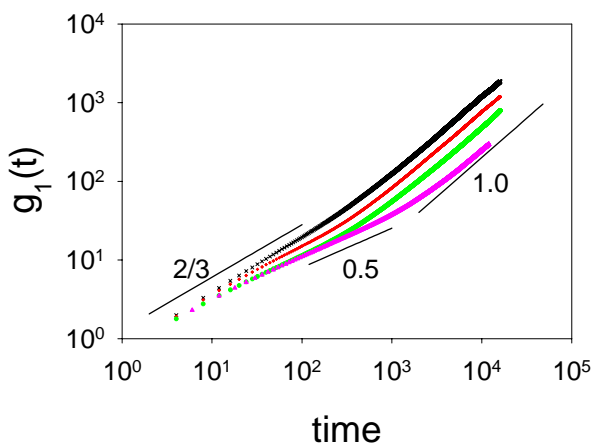

**Fig. 11**: The monomer diffusive motion. Data from top to bottom are for N =50 at $\phi$=0.05, N=50 at $\phi$=0.3, N=50 at $\phi$=0.7 and N=100 at $\phi$=0.7. The black straightlines indicate the scaling exponents expected for the Zimm model (2/3), Rouse model (0.5), and the long time limit regime (1.0).

**3.2.D Relaxation mode analysis**: As a further investigation of equilibrium dynamics in polymer solutions, we also monitored the dependence of the relaxation time $\tau_p$ with respect to mode index $p$ as the polymer volume fraction is increased. As both the excluded volume interaction and the hydrodynamic interaction becomes screened, the exponent, $n$, in the power law, $\tau_p \sim p^{-n}$, is expected to vary from $n = 3\nu \approx 1.77$ (using $\nu \approx 0.59$) for a Zimm-like chain to $n = 2.0$ for a Rouse-chain with increasing polymer volume fraction. As shown in Fig. 12(a), analysis of the first ten modes for two polymer solutions with $\phi = 0.05$ and $\phi = 0.5$, both with $N=100$ lead to $n = 1.86 \pm 0.03$ for $\phi = 0.05$, and $n = 1.93 \pm 0.02$ for $\phi = 0.5$ (the later correspond to $\phi/\phi^*$=6.2). Although the change in the exponent, $n$, is small, it reflects the screening of both the hydrodynamic interactions and excluded volume interactions as the polymer volume fraction is increased. The dependence of $\tau_p$ on the Rouse amplitude, $R_p$, is shown in Fig. 12(b). In the dilute solution limit, $\phi = 0.05$, we observe that $\tau_p \sim R_p^{3.05 \pm 0.06}$, as reported earlier in section 3.1.C. However, as the polymer volume fraction is increased to $\phi = 0.5$, the exponent increases to $3.59 \pm 0.05$. We recall that the Rouse model predicts that $\tau_p \sim R_p^{2+1/\nu}$. At $\phi = 0.5$, we found the Flory exponent $\nu \approx 0.54$. Therefore, we predict $\tau_p \sim R_p^{3.85}$ for $\phi = 0.5$ if the chains behave as Rouse chains. The fact that the numerically found exponent lies between those predicted by the Zimm model and the Rouse model, $3.0 < 3.59 < 3.85$, implies that at $\phi = 0.5$, hydrodynamic interactions are partially screened.

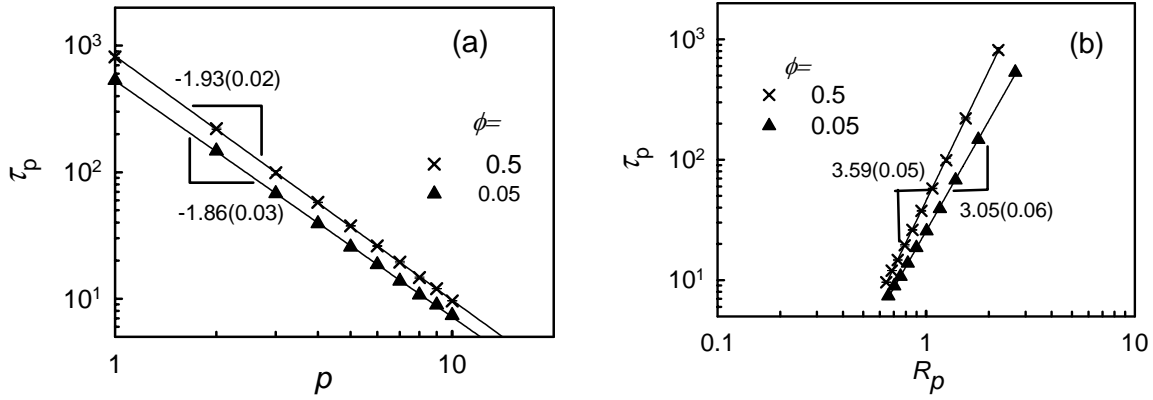

**Fig. 12:** Plot of $p$th mode relaxation time $\tau_p$ versus (a) mode index $p$, and (b) mode amplitude $R_p$ at two different volume fractions, $\phi=0.05$ and 0.5 for a polymer solution with a chain length $N=100$. The solid lines are the power law fits of the data.

## 3.3 Dynamics of polymers in dilute solutions with varying Schmidt number

Finally, we present results based on DPD simulations of polymer dynamics in dilute solutions with varying Schmidt number, defined as $\mathrm{Sc} \equiv \mu_k/D_s$. There are several ways to achieve a large Schmidt number in DPD simulations. One way is to use a large cutoff radius in DPD simulation, but this increases the CPU time significantly. Another way is to reduce system temperature which leads to a reduction in diffusion coefficient, and hence an increase in the Schmidt number.[31] Groot and Warren gave approximate expressions for the diffusion coefficient

$$D_s \approx \frac{45 m k_B T}{2\pi \gamma \rho r_c^3}, \qquad (20)$$

and kinematic viscosity

$$\mu_k \approx \frac{D_s}{2} + \frac{2\pi \gamma \rho r_c^5}{1575 m}, \qquad (21)$$

of a DPD fluid. In Eqs. (20) and (21), $\gamma$ is the prefactor of the dissipative force and is related to parameter $\sigma$ through fluctuation-dissipation theorem $\sigma^2 = 2\gamma k_B T$. The kinematic viscosity of a DPD fluid is dominated by the second term in Eq. (21). Therefore, the Schmidt number is approximately given by

$$\mathrm{Sc} \approx \frac{1}{2} + \frac{(2\pi \gamma \rho r_c^4)^2}{70875 m^2 k_B T}, \qquad (22)$$

In our simulation, we set $\sigma$ at a constant value, $\sigma = 3.0(\varepsilon^3 m / r_c^2)^{1/4}$, and varied temperature $T$ for a dilute solution of polymers with chain length $N = 60$. Clearly, according to Eq. (22), when $T$ increases, the Schmidt number would decrease whereas when $T$ decreases, the Schmidt number would increase.

We investigated the dynamic properties of polymer solutions with temperature, $k_B T / \varepsilon$, varying between 0.4 and 2.1. Fig. 13 shows the scaling dependence of $\tau_p \sim R_p^n$ at three temperatures, $k_B T / \varepsilon = 0.4$, 1.0, and 2.1. The exponent $n$ is found to be $2.84 \pm 0.05$, $3.05 \pm 0.05$, and $3.33 \pm 0.05$, respectively. As temperature increases above 1.0, the Schmidt number is further reduced. The observed exponent is seen to increase toward the value expected from the Rouse model. This implies that at high temperatures with a further reduction in the Schmidt number, the hydrodynamic interaction within the chain is not fully developed. This leads to an increased exponent toward the value expected from the Rouse model. However, at $k_B T / \varepsilon = 2.1$, the observed exponent $3.33 \pm 0.05$ is still smaller than the Rouse exponent, 3.69. The polymer chain exhibit partial draining at this temperature.

The decrease of the exponent at lower $T$ was puzzling to us. In principle, the increase of the Schmidt number as the temperature lowers helps the development of hydrodynamic interaction. We would therefore expect the results in better agreement with the Zimm model. The physical meaning of the exponent being smaller than the Zimm value is not clear to us, but we note the exponent changes in the opposite direction of the Rouse model.

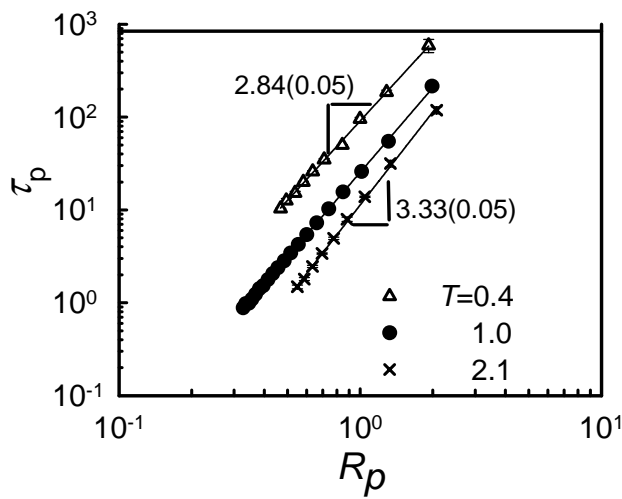

**Fig. 13** Log-log plot of $\tau_p$ versus $R_p$ at different temperatures: $T = 0.4$ (Δ); $T=1.0$ (●) and $T=2.1$ (×). The volume fraction of the polymers is at $\phi=0.004$ with a chain length $N =60$. The solid lines represent power law fits in the form of $\tau_p \sim R_p^n$.

## 4 Summary and Conclusions

In this study, we investigated in detail the dynamics of both dilute and concentrated polymer solutions in good solvent using dissipative particle dynamics simulation. In particular, we compared our results with those predicted from the two well known theoretical models, the Zimm model, which accounts for hydrodynamic effects, and the Rouse model, which ignores hydrodynamics. The comparison aims to reveal the development of the hydrodynamic interaction within polymer chains simulated with the DPD approach. From our extensive analysis, we conclude the following:

1) In dilute solution, the chain simulated with the standard DPD approach possess an excluded volume interaction when the solvent/polymer interaction $a_{ps}$ is kept the same as $a_{pp}$ and $a_{ss}$. There is no need to introduce additional hard-core or Lennard-Jones interactions between polymer beads to account for the excluded volume interaction, although the use of such additional force may help to preserve the topological constraint.

2) The polymer dynamics in dilute solution obeys the Zimm model, not the Rouse model at $k_BT/\varepsilon=1.0$. The clear evidence comes from the dependence of $\tau \sim R_g^{3.0\pm0.1}$ and $\tau_p \sim R_p^{3.05\pm0.06}$. This implies that the hydrodynamic interaction is reasonably developed within the polymer chain at common DPD simulation conditions. When the Schmidt number is further decreased (in this study, through increasing *T*), the deviation of polymer dynamics from the Zimm behavior towards Rouse behavior is observed. This implies that a low Schmidt number in DPD simulation may lead to an underdevelopment of hydrodynamic interaction within the chains. However, this occurs not at common conditions employed in DPD simulation, but occurs at a higher *T* with even lower Schmidt number.

3) As the polymer volume fraction increases, the hydrodynamic interaction and the excluded volume interaction are both screened. The screening process follows the theoretical predictions reasonably well. The use of soft interaction between polymer beads and a low Schmidt number have not produced noticeable problems for the simulated dynamics, except that the entanglement effect is not captured in the simulations.

Our study suggests that the concerns that because the Schmidt number in DPD simulation is just on the order of one and therefore DPD is not suitable to simulate polymer dynamics in dilute solution, are not well established. DPD is a coarse-grained, particle-based simulation method. In DPD simulation, each *dpd*-particle represents a packet of fluid. For typical fluids, the Schmidt number is defined as the kinematic viscosity over the self-diffusion coefficient, $Sc \equiv \mu_k/D_s$, where $D_s$ is the self diffusion coefficient of individual molecules. In a coarse-grained method such as with DPD, the self-diffusion coefficient of a *dpd*-particle does not correspond to the self-diffusion coefficient of individual solvent molecules. Therefore, as suggested by Peters, the Schmidt number is an ill-defined quantity in a coarse-grained approach.[47] We note that in MD simulations of bead-spring models of polymer chains in explicit solvent particles, the Schmidt number thus calculated was also small. Polson and Gallant[14], for example, gave a Schmidt number Sc=27.2 and 73.0, at a number density $\rho=0.8\sigma^{-3}$ and $0.9\sigma^{-3}$ respectively. Several earlier MD simulations in explicit solvent, which confirmed the conformity of simulated polymer dynamics to the Zimm model, were performed at similar conditions by Polson and Gallant.[14] From the simulation point of view, Peters suggested that one would like to use Sc of order of one. There is no need to simulate models with "realistically" large Sc ≈ 1000 in order to capture the hydrodynamic interactions. Furthermore, a model with such a large Scmidt number may be prohibitively expensive in terms of CPU time.

Our study suggests that the hydrodynamic interaction (HI) is developed within the size of chains during its own relaxation time under typical DPD simulations. We think that polymer beads within a chain are spatially close enough such that the development of HI can be attained.

This however may not preclude underdevelopment of HI between particles separated over much larger spatial distance. For example, the development of HI between polymer chains with channel walls may still have problems. Recently, Fan et al[48] has shown that the velocity profile of a DPD fluid is underdeveloped when flowing through a channel with abrupt contraction. The DPD particles do not feel the presence of the walls until they are very close to the walls. In future, we will present our DPD simulations of dilute polymer solutions flowing through a channel. There, chain migration pattern observed in DPD simulations is seen to be influenced by the Schmidt number of the simulated DPD solvent.

## Acknowledgements

We would like to thank the Petroleum Research Fund from the American Chemical Society, and a grant from The University of Memphis Faculty Research Grant. The later support does not necessarily imply endorsement by the university of research conclusions. We also like to thank Prof. James Polson for useful discussions.

# References

(1) Rubinstein, M., Colby, R. H.; Oxford University Press: New York, 2003.
(2) de Gennes, P. G. *Scaling Concepts of Polymer Physics*; Cornell University Press: Ithaca, 1979.
(3) Doi, M., Edwards, S.F. *The Theory of Polymer Dynamics*; Clarendon: Oxford, 1986.
(4) Rouse, P. E. J. *J. Chem. Phys.* **1953**, *21*, 1272-1280.
(5) Zimm, B. H. *J. Chem. Phys.* **1956**, *24*, 269-278.
(6) Smith, D. E., Perkins, T.T., Chu, S. *Macromolecules* **1996**, *29*, 1372-1373.
(7) Hur, J. S., Shaqfeh, E.S.G., Babcock, H.P., Chu, S. *Phys. Rev. E.* **2002**, *66*, 011915.
(8) Hur, J. S., Shaqfeh, E.S.G., Larson, R.G. *J. Rheol.* **2000**, *44*, 713-741.
(9) Smith, D. E., Babcock, H.P., Chu, S. *Science* **1999**, *283*, 1724-1727.
(10) Kaznessis, Y. N., Hill, D. A., Maginn, E.J. *J. Chem. Phys.* **1998**, *109*, 5078-5088.
(11) Dünweg, B.; Kremer, K. *Phys. Rev. Lett.* **1991**, *66*, 2996.
(12) Pierleoni, C. R., J.-P. *Phys. Rev. Lett.* **1992**, *61*, 2992.
(13) Dünweg, B., Kremer, K. *J. Chem. Phys.* **1993**, *99*, 6983-6997.
(14) Polson, J. M.; Gallant, J. P. *J. Chem. Phys.* **2006**, *124*, 184905.
(15) Jendrejack, R. M., de Pablo, J.J., Graham, M.D. *J. Chem. Phys.* **2002**, *116*, 7752-7759.
(16) Jendrejack, R. M., Dimalanta, E.T., Schwartz, D.C., Graham, M.D., de Pablo, J.J *Phys. Rev. Lett.* **2003**, *91*, 038102.
(17) Jendrejack, R. M., Schwartz, D.C., de Pablob, J.J. *J. Chem Phys.* **2003**, *119*, 1165-1173.
(18) Jendrejack, R. M., Schwartz, D.C., de Pablob, J.J., Graham, M.D. *J. Chem. Phys.* **2004**, *120*, 2513-2529.
(19) Ahlrichs, P., Dunweg, B. *J. Chem. Phys.* **1999**, *111*, 8225-8239.
(20) Usta, O. B., Butler, J. E., Ladda, A. J. C. *Phys. Fluids* **2006**, *18*, 031703.
(21) Usta, O. B.; Ladd, A. J. C.; Butler, J. E. *J. Chem. Phys.* **2005**, *122*, 094902.
(22) Hoogerbrugge, P. J., Koelman, J.M.V.A. *Europhys. Lett.* **1992**, *19*, 155-160.
(23) Español, P., Warren, P. *Europhys. Lett.* **1995**, *30*, 191-196.
(24) Español, P. *Europhys. Lett.* **1997**, *40*, 631.
(25) Ripoll, M.; Ernst, M. H.; Espanol, P. *J. Chem. Phys.* **2001**, *115*, 7271-7284.
(26) Groot, R. D., Madden, T.J., Tildesley, D.J. *J. Chem. Phys.* **1999**, *110*, 9739-9748.
(27) Laradji, M., Hore, M.J.A. *J. Chem. Phys.* **2004**, *121*, 10641.
(28) Laradji, M., Kumar, P.B.S. *Phys. Rev. Lett.* **2004**, *93*, 198105.
(29) Wijmans, C. M.; Smit, B. *Macromolecules* **2002**, *35*, 7138-7148.
(30) Huang, J.; Wang, Y.; Laradji, M. *Macromolecules* **2006**, *39*, 5546.
(31) Groot, R. D., Warren, P.B. *J. Chem. Phys.* **1997**, *107*, 4423-4435.
(32) Pan, G.; Manke, C. W. *Int. J. Mod. Phys. B* **2003**, *17*, 231-235.
(33) Kong, Y., Manke, C. W., Madden,W. G. *J. Chem. Phys.* **1997**, *107*, 592-602.
(34) Pan, G.; Manke, C. W. *J. Rheol.* **2002**, *46*, 1221-1237.
(35) Spenley, N. A. *Europhys. Lett.* **2000**, *49*, 534-540.
(36) Pagonabarraga, I., Hagen, M.H.J., Frenkel D. *Europhys. Lett.* **1998**, *42*, 377.
(37) Milchev, A., Binder, K. *J. Chem. Phys.* **1997**, *106*, 1978.
(38) Milchev, A., Rostiashvili V. G., Vilgis, T. A. *Europhys. Lett.* **2004**, *68*, 384-390.
(39) Schlijper, A. G., Hoogerbrugge, P. J., Mankea, C. W. *J. Rheol.* **1995**, *39*, 567-579.
(40) Symeonidis, V., Karniadakis,G.E., Caswell, B. *Phys. Rev. Lett.* **2005**, *95*, 076001.
(41) Dunweg, B.; Reith, D.; Steinhauser, M.; Kremer, K. *J. Chem. Phys.* **2002**, *117*, 914-924.
(42) Kremer, K., Grest, G.S. *J. Chem. Phys.* **1990**, *92*, 5057-5086.
(43) Wang, Y., Teraoka, I. *Macromolecules* **2000**, *33*, 3478.
(44) Muthukumar, M. *Macromolecules* **1984**, *17*, 971-973.

(45) Muthukumar, M., Freed, K.F. *Macromolecules* **1978**, *11*, 843.
(46) Patel, S. S., Takahashi, K.M. *Macromolecules* **1992**, *25*, 4382.
(47) Peters, E. A. J. F. *Europhys. Lett.* **2004**, *66*, 311-317.
(48) Fan, X.; Phan-Thien, N.; Chen, S.; Wu, X.; Ng, T. Y. *Phys. Fluids.* **2006**, *18*, 063102.